%% file: summary.tex
\documentclass[sigconf, authorversion]{acmart}

\setcopyright{none}

\input{additional-packages}

\input{custom-definitions}
\input{acronyms}

\graphicspath{ {./figures/} }

\acmDOI{}

\acmConference[NetMob '23]{Primary Conference on the Analysis of Mobile Phone Datasets in Social, Urban, Societal and Industrial Problems}{October 04--06,
  2023}{Madrid, Spain}

\begin{document}   

\title{Maintaining App Services in Disrupted Cities: A Crisis and Resilience Evaluation Tool}
\author{Leon W\"ursching}
\orcid{0000-0003-2648-6507}
\affiliation{%
  \department{Secure Mobile Networking Lab}
  \institution{Technical University of Darmstadt}
  \city{Darmstadt}
  \country{Germany}
}
\email{lwuersching@seemoo.tu-darmstadt.de}
\author{Matthias Hollick}
\orcid{0000-0002-9163-5989}
\affiliation{%
  \department{Secure Mobile Networking Lab}
  \institution{Technical University of Darmstadt}
  \city{Darmstadt}
  \country{Germany}
}
\email{mhollick@seemoo.tu-darmstadt.de}
\settopmatter{printacmref=false}
\renewcommand{\shortauthors}{W\"ursching et al.}

\begin{abstract}
Disaster scenarios can disconnect entire cities from the \gls{cn}, isolating \glspl{bs} and disrupting the Internet connection of app services for many users.
Such a disruption is particularly disastrous when it affects \emph{critical app services} such as communication, information, and navigation.
Deploying local app servers at the network edge can solve this issue but leaves \glspl{mno} faced with design decisions regarding the criticality of traffic flows, the \gls{bs} topology, and the app server deployment.
\textbf{We present the Crisis and Resilience Evaluation Tool (\caret) for crisis-mode \glspl{ran}}, enabling \glspl{mno} to make informed decisions about a city's \gls{ran} configuration based on real-world data of the NetMob23 dataset.
\end{abstract}

\maketitle

%%%%%%%%%%%%%%%%%%%%%%%%%%%%%%%%%%%%%%%%%%%%%%%%%%%%%%%%%%%%%%%%%%%%%%%%%%%%%%%%%%%%%%%%%%%%%%%%%%%%
\section{Scenario and Goal}\label{sec:scenario}
%%%%%%%%%%%%%%%%%%%%%%%%%%%%%%%%%%%%%%%%%%%%%%%%%%%%%%%%%%%%%%%%%%%%%%%%%%%%%%%%%%%%%%%%%%%%%%%%%%%%

\begin{figure}
  \centering
  \includegraphics[width=\columnwidth]{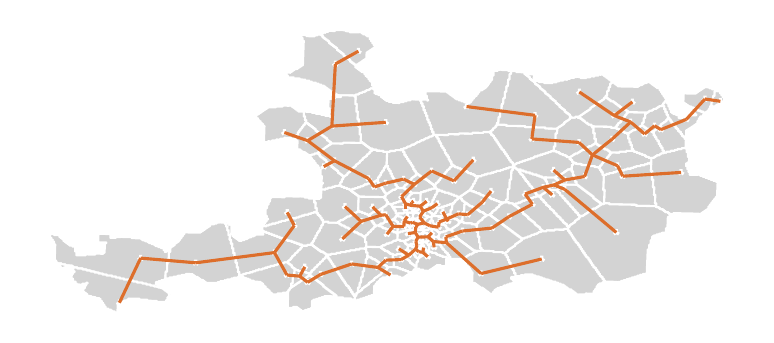}
  \caption{The RAN of Saint-Etienne in crisis mode showcasing each BS's coverage (gray tiles) and the minimum set of inter-BS links connecting all BSs (orange lines).}
  \Description{The map of Saint-Etienne containing all tiles of the NetMob23 spatial data set. The tiles are partitioned per base station and indicated as gray areas. Also depicted are the links of the inter-BS network, where each link is shown as an orange line. The chosen network topology is a minimum spanning tree.}
  \label{fig:scenario}
\end{figure}

We assume a disaster scenario where a city is disconnected from the \gls{cn}, but a set of \glspl{bs} is still available.
\textbf{Our goal is to maintain app services for the users of a disrupted city by deploying local app services on the network edge}.
In this context, we focus on the user plane, so our scenario assumes that the \gls{mno} has already attended to the availability of the control plane and the necessary \gls{cn} functions.
Any approaches to reconnect the city to the \gls{cn} %\cite{minh2016site}
or the Internet %\cite{minh2014fly}
are outside the scope of this work as we provide a solution within the disrupted city.

Starting from today's cellular networks, many decisions have to be made so that the app services can be maintained in disrupted cities:
Our vision (cf. \autoref{fig:scenario}) is that the users remain connected to the closest \gls{bs}, app services are deployed on local edge servers, and \glspl{bs} route traffic between users and app services via wireless links.

%%%%%%%%%%%%%%%%%%%%%%%%%%%%%%%%%%%%%%%%%%%%%%%%%%%%%%%%%%%%%%%%%%%%%%%%%%%%%%%%%%%%%%%%%%%%%%%%%%%%
\section{Contribution}
%%%%%%%%%%%%%%%%%%%%%%%%%%%%%%%%%%%%%%%%%%%%%%%%%%%%%%%%%%%%%%%%%%%%%%%%%%%%%%%%%%%%%%%%%%%%%%%%%%%%

\begin{figure}
  \centering
  \includegraphics[scale=1]{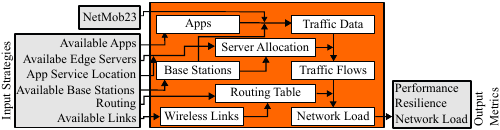}
  \caption{System model of \caret, including input parameters, the data flow, and the resulting output metrics.}
  \Description{The model consists of four components: On the left, there are the input components "NetMob23" and "Input Strategies", in the center, there is the CARET component, and on the right there is the "Output Metrics" component. Arrows indicate, how the input arguments are processed to determine the available apps, base stations, and wireless links. From these, the server allocation and routing table is generated, which supports the transformation of the provided traffic data into traffic flows into network load.}
  \label{fig:system-model}
\end{figure}

For the \gls{ran} to support such a crisis scenario, the \gls{mno} has to make many decisions, e.g., about the available infrastructure and supported traffic.
\textbf{Our contribution is the evaluation tool \caret for crisis-relevant decisions with real-life mobile traffic data from the NetMob23 data set}.
\caret (cf. \autoref{fig:system-model}) expects mobile traffic data and the following crisis decisions as input:

\begin{itemize}
  \hlitem{Apps} Which apps should supported in crisis mode?
  \hlitem{Base Stations} Which \glspl{bs} are available in the city
  \hlitem{Edge Servers} On which \glspl{bs} can edge servers be provided? 
  \hlitem{App Servers} Where are the app services deployed?
  \hlitem{Links} Which links are available to connect \glspl{bs}?
  \hlitem{Routing} How is traffic routed through the city?
\end{itemize}

It then evaluates how well the given decisions perform for the provided traffic data and returns the following evaluation metrics:

\begin{itemize}
  \hlitem{Resilience} Which fraction of the traffic can be served?
  \hlitem{Performance} How much traffic is routed wirelessly?
  \hlitem{Load} What is the overall traffic load in the city?
\end{itemize}

%%%%%%%%%%%%%%%%%%%%%%%%%%%%%%%%%%%%%%%%%%%%%%%%%%%%%%%%%%%%%%%%%%%%%%%%%%%%%%%%%%%%%%%%%%%%%%%%%%%%
\section{Data Preparation}
%%%%%%%%%%%%%%%%%%%%%%%%%%%%%%%%%%%%%%%%%%%%%%%%%%%%%%%%%%%%%%%%%%%%%%%%%%%%%%%%%%%%%%%%%%%%%%%%%%%%

Each input parameter to \caret \cite{seemoo2023caret} is either a concrete \textit{configuration}, e.g., the set of available \glspl{bs}, or a \textit{strategy}, such as \strategy{HIGH TRAFFIC 80}, which uses the provided traffic data to identify the \glspl{bs} with the highest traffic volume and make 80\% of them available.

\caret's required traffic data format is very similar to that of the NetMob23 data set \cite{netmob2023dataset}, with the following differences:
On the spatial dimension, we consider \gls{bs}-wise instead of tile-wise traffic utilizing the \gls{bs} information available at Cartoradio \cite{cartoradio2023stations}.
On the temporal dimension, we deviate from the NetMob23 directory and file structure to obtain a file for each time slot, facilitating parallel evaluation.
\textbf{We provide a conversion tool for the NetMob23 data set to be compatible with \caret.}

\begin{figure}[t]
  \centering
  \includegraphics[width=\columnwidth]{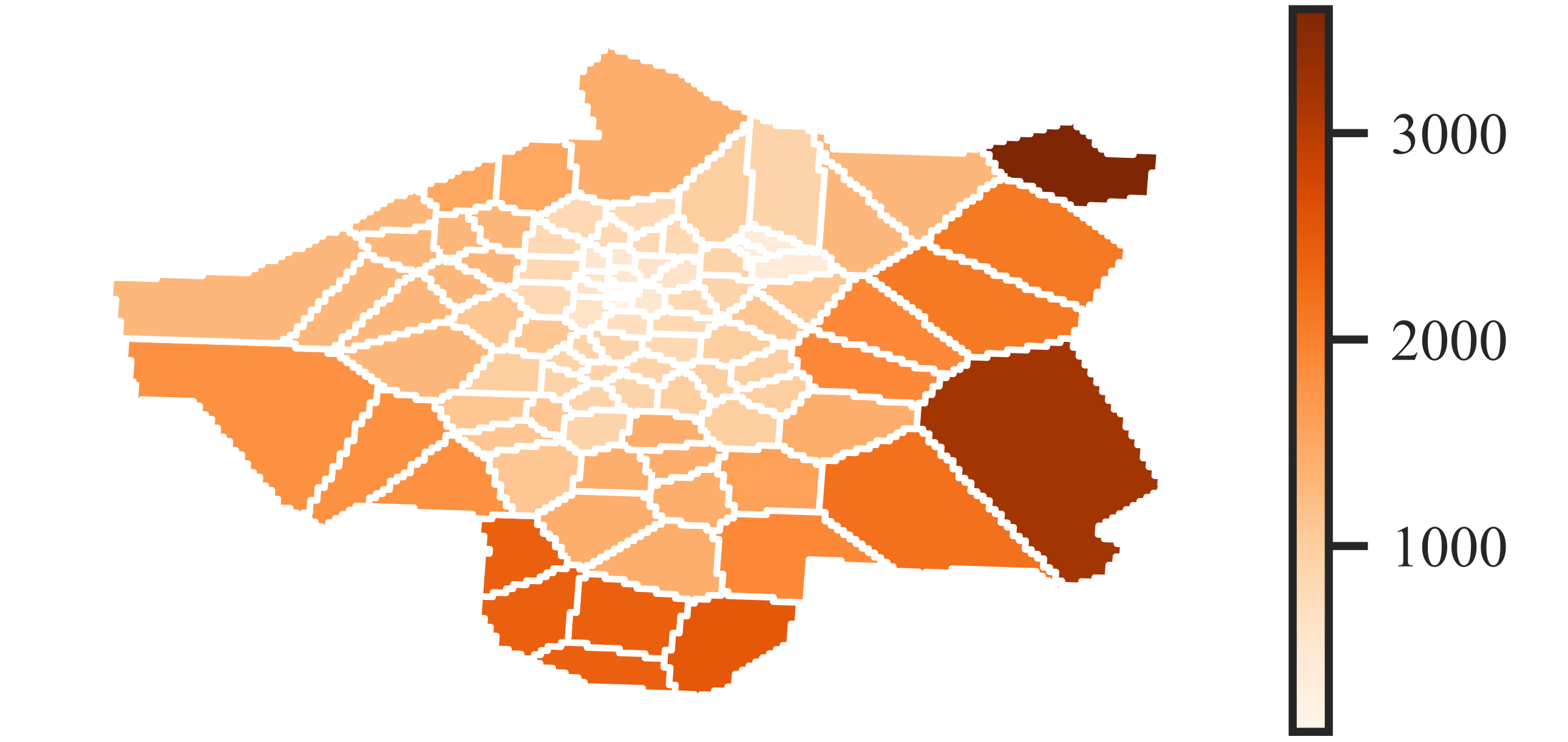}
  \caption{Minimum wireless link range [m] required for the BSs in the city of Nancy to connect to the inter-BS network.}
  \Description{The map of Nancy containing all tiles of the NetMob23 spatial data set. The tiles are partitioned per base station and indicated as individual areas. The areas are colored to depict the minimum link range required to connect them to the inter-BS network. The required link range is lowest in the city center and increases with distance to the city center.}
  \label{fig:connectivity}
\end{figure}

\begin{figure}[t]
  \centering
  \includegraphics[scale=1]{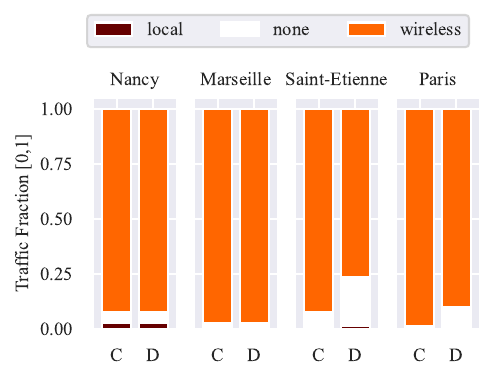}
  \caption{
    RAN performance for a connectivity of 85\%.
    Depicted is the traffic fraction that is handled locally, routed wirelessly, and cannot be served for the app service strategies \strategy{CENTRAL} (C) and \strategy{DECENTRAL} (D).
    }
  \Description{Depicted is a stacked bar chart with two bars (central and decentral strategy) for each of the four cities (Nancy, Marseille, Saint-Etienne, and Paris). The stacked bar charts depict the traffic fraction that is non-servicable, routed wirelessly, or handled locally.}
  \label{fig:performance}
\end{figure}

%%%%%%%%%%%%%%%%%%%%%%%%%%%%%%%%%%%%%%%%%%%%%%%%%%%%%%%%%%%%%%%%%%%%%%%%%%%%%%%%%%%%%%%%%%%%%%%%%%%%
\section{Evaluation}
%%%%%%%%%%%%%%%%%%%%%%%%%%%%%%%%%%%%%%%%%%%%%%%%%%%%%%%%%%%%%%%%%%%%%%%%%%%%%%%%%%%%%%%%%%%%%%%%%%%%

As depicted in \autoref{fig:system-model}, we filter the input traffic data by available apps and \glspl{bs}.
For each app service, we select one edge-capable \gls{bs} to host the local app service.
This enables us to generate traffic flows while considering two traffic profiles:
The uplink traffic of each \gls{bs} is partitioned per app and routed to the corresponding app service, and the downlink traffic is routed on the reverse path.
Each traffic flow is routed through the city based on the routing table and the available links.
Then, we compute the network load by accumulating the individual link loads.

%%%%%%%%%%%%%%%%%%%%%%%%%%%%%%%%%%%%%%%%%%%%%%%%%%%%%%%%%%%%%%%%%%%%%%%%%%%%%%%%%%%%%%%%%%%%%%%%%%%%
\section{Results}
%%%%%%%%%%%%%%%%%%%%%%%%%%%%%%%%%%%%%%%%%%%%%%%%%%%%%%%%%%%%%%%%%%%%%%%%%%%%%%%%%%%%%%%%%%%%%%%%%%%%
\caret is applicable to the entire NetMob23 data set \cite{netmob2023dataset}.
However, we limit our reporting to the following cities to showcase different combinations of size and population density:
Nancy (small and sparse), Marseille (small and dense), Saint-Etienne (large and sparse), and Paris (large and dense).
For this section, we assume the role of an \gls{mno} configuring a city's \gls{ran} in response to a crisis.

\subsection{Link Range vs. Energy Consumption} %%%%%%%%%%%%%%%%%%%%%%%%%%%%%%%%%%%%%%%%%%%%%%%

Wireless link establishment in ad-hoc networks is a well-researched problem, and an optimal configuration highly depends on individual infrastructure characteristics.
Therefore, we skip concrete configurations and instead showcase \caret's functionality with a simple model where each \gls{bs} connects to all neighboring \glspl{bs} in a radius of $r$ meters.
The \gls{mno} has to trade off service coverage and energy consumption:
A larger radius leads to higher inter-\gls{bs} connectivity but causes higher energy consumption.
\autoref{fig:connectivity} shows the minimum link range required for each \gls{bs} in Nancy to connect to the inter-\gls{bs} network.
The required radius for a connectivity of 85\% is \SI{1200}{\meter} in Paris, \SI{1400}{\meter} in Marseille, \SI{1500}{\meter} in Nancy, and \SI{4300}{\meter} in Saint-Etienne.
The \gls{mno} can use \caret to evaluate different link establishment algorithms and decide which link range is required to achieve the desired connectivity.

\subsection{Central vs. Decentral Service Deployment} %%%%%%%%%%%%%%%%%%%%%%%%%%%%%%%%%%%%%%%%%%%%%%

Deploying app services on the local network edge is a non-trivial allocation problem, so we showcase two simple deployment strategies:
Deploying all app services on one edge server with the most traffic (\strategy{CENTRAL}) and deploying each service at the edge server where there is the most app-specific traffic (\strategy{DECENTRAL}).
For \autoref{fig:performance}, we evaluated the traffic data of May 31, 2019, with the following strategies:
All apps and all \glspl{bs} are available, all \glspl{bs} are edge-enabled, and each city's link range is set to achieve 85\% connectivity with minimum distance routing.
The fraction of locally handled traffic n dense cities is negligible, and non-serviceable traffic is lower than in sparse cities.
While both strategies perform similarly in small cities, the \strategy{CENTRAL} strategy dominates in large cities.
The high fraction of non-serviceable traffic for the \strategy{DECENTRAL} strategy in Saint-Etienne is caused by app services deployed at one of the 15\% disconnected \glspl{bs}.
\caret enables \glspl{mno} to evaluate the performance of deployment strategies for local app services based on recorded data, e.g., the NetMob23 data set \cite{netmob2023dataset}.

%%%%%%%%%%%%%%%%%%%%%%%%%%%%%%%%%%%%%%%%%%%%%%%%%%%%%%%%%%%%%%%%%%%%%%%%%%%%%%%%%%%%%%%%%%%%%%%%%%%%
\section{Conclusion and Future Work}
%%%%%%%%%%%%%%%%%%%%%%%%%%%%%%%%%%%%%%%%%%%%%%%%%%%%%%%%%%%%%%%%%%%%%%%%%%%%%%%%%%%%%%%%%%%%%%%%%%%%

The goal of this work is to maintain app services in disrupted cities by deploying app services on the local network edge.
To this end, we introduce the Crisis and Resilience Evaluation Tool (\caret) \cite{seemoo2023caret}, supporting \glspl{mno} to make informed decisions for the configuration of crisis-mode \glspl{ran}.
Furthermore, we provide a conversion tool that enables \glspl{mno} to use \caret with the NetMob23 data set.

For future work, we envision a cellular network that can transition into crisis mode, as described in Section\xspace\ref{sec:scenario}.
We will continue the development of \caret to support \glspl{mno} in this endeavor.

\bibliographystyle{ACM-Reference-Format}
\bibliography{bibliography}

\end{document}

%% file: additional-packages.tex
\usepackage{cleveref}
\usepackage{colortbl}
\usepackage{csquotes}
\usepackage{float}
\usepackage[acronyms]{glossaries}
\glsdisablehyper
\usepackage{hyperref}
\usepackage{multirow}
\usepackage{pifont}
\usepackage{siunitx}
\usepackage{subcaption}
\usepackage{tabularx}
\usepackage{threeparttable}
\usepackage{xurl}
\usepackage{xspace}

%% file: custom-definitions.tex
\newcommand{\hlitem}[1]{\item\textbf{#1}.}

\newcommand{\caret}{CARET\xspace}

\newcommand{\strategy}{\texttt}

%% file: acronyms.tex
\newacronym[]{bs}{BS}{base station}
\newacronym[]{cp}{C-Plane}{control plane}
\newacronym[]{cn}{CN}{core network}
\newacronym[]{csv}{CSV}{comma-separated values}
\newacronym[]{mno}{MNO}{mobile network operator}
\newacronym[]{mst}{MST}{minimum spanning tree}
\newacronym[]{ran}{RAN}{radio access network}
\newacronym[]{up}{U-Plane}{user plane}